%% file: 0main.tex
\documentclass[sigconf, 9pt]{acmart}

\usepackage{amsmath,amssymb,amsfonts}
\usepackage{graphicx}
\usepackage{textcomp}
\usepackage{xcolor}
\usepackage{soul}
\usepackage{algorithm}
\usepackage{wrapfig}
\usepackage{multirow}
\usepackage{subcaption}
\usepackage{threeparttable}
\usepackage{colortbl}
\definecolor{mygray}{gray}{.85}
\usepackage{enumitem}
\usepackage{algorithm, algpseudocode}

\AtBeginDocument{%
  }

\copyrightyear{2025}
\acmYear{2025}
\setcopyright{cc}
\setcctype{by}
\acmConference[BUILDSYS '25]{The 12th ACM International Conference on Systems for Energy-Efficient Buildings, Cities, and Transportation}{November 19--21, 2025}{Golden, CO, USA}
\acmBooktitle{The 12th ACM International Conference on Systems for Energy-Efficient Buildings, Cities, and Transportation (BUILDSYS '25), November 19--21, 2025, Golden, CO, USA}
\acmDOI{10.1145/3736425.3770114}
\acmISBN{979-8-4007-1945-5/2025/11}

\makeatother

\begin{document}

\title[Enhancing Shared Micromobility Efficiency through Minimal Autonomous Vehicle Deployment]{Small Fleet, Big Impact: Enhancing Shared Micromobility Efficiency through Minimal Autonomous Vehicle Deployment}

\author{Heng Tan}
\affiliation{%
    \institution{Lehigh University}
  \city{Bethlehem}
  \country{USA}
  }
\email{het221@lehigh.edu
}
\author{Hua Yan}
\affiliation{%
  \institution{Lehigh University}
  \city{Bethlehem}
  \country{USA}}
\email{huy222@lehigh.edu
}

\author{Lucas Yang}
\affiliation{%
  \institution{Parkland High School}
  \city{Allentown}
  \country{USA}}
\email{yanglucas2028@gmail.com
}

\author{Yu Yang}
\affiliation{%
  \institution{Lehigh University}
  \city{Bethlehem}
  \country{USA}}
\email{yuyang@lehigh.edu
}

\renewcommand{\shortauthors}{H. Tan, H. Yan, L. Yang, Y. Yang}



\begin{abstract}
Shared micromobility systems, such as electric scooters and bikes, have gained widespread popularity as sustainable alternatives to traditional transportation modes. However, these systems face persistent challenges due to spatio-temporal demand fluctuations, often resulting in a mismatch between vehicle supply and user demand. Existing shared micromobility vehicle scheduling methods typically redistribute vehicles once or twice per day, which makes them vulnerable to performance degradation under atypical conditions. In this work, we design to augment existing micromobility scheduling methods by integrating a small number of autonomous shared micromobility vehicles (ASMVs), which possess self-rebalancing capabilities to dynamically adapt to real-time demand. Specifically, we introduce \textbf{SMART}, a hierarchical reinforcement learning framework that jointly optimizes high-level initial deployment and low-level real-time rebalancing for ASMVs. We evaluate our framework based on real-world e-scooter usage data from Chicago. Our experiment results show that our framework is highly effective and possesses strong generalization capability, allowing it to seamlessly integrate with existing vehicle scheduling methods and significantly enhance overall micromobility service performance.

\end{abstract}

\maketitle

\input{1Introduction}

\input{3Motivation}

\input{8ProblemDefinition}

\input{4Method}

\input{5Evaluation}

\input{2RelatedWork}

\input{7Conclusion}

\newpage
\bibliographystyle{ACM-Reference-Format}
\bibliography{refs}

\end{document}

%% file: 1Introduction.tex
\section{Introduction}

\textbf{Background}: Shared micromobility (e.g., e-scooters and bikes), as an alternative urban transport way to conventional cars, has gained worldwide popularity in recent years. For instance, in the United States, more than 157 million trips were made using shared bikes and scooters in 2023, with e-scooters accounting for nearly 50$\%$ of the total~\cite{NACTO2023micromobility}. This rapid adoption is largely due to their convenience, lower environmental footprint, and ability to alleviate traffic congestion in dense urban areas~\cite{zhong2023rlife,zhong2024adatrans}. However, the rapid growth of shared micromobility systems has introduced significant management challenges. Among them, one of the most critical issues is the persistent spatio-temporal imbalance between vehicle supply and user demand~\cite{tan2023joint}, which not only reduces user satisfaction but also increases unnecessary operational costs. In this work, we use e-scooters as an example to study the problem of shared micromobility vehicle scheduling.

\textbf{State-of-The-Art (SoTA) and Limitations}: Many scheduling methods for shared micromobility vehicles have been designed in recent years~\cite{tan2024robust,wu2024fleet,he2022socially,zhao2024urban,wang2021record,jaller2021dock,yang2024mallight}. 
They generally decide how to schedule vehicles based on the current spatial vehicle distribution and (or) predicted user demand before the next scheduling time.
To optimize scheduling decisions, they typically employ either mixed-integer programming methods~\cite{yuan2019p,guo2020vehicle,zhao2018integrated} or sequential decision-making frameworks such as reinforcement learning~\cite{he2023robust,tan2023joint,tan2024robust} to learn policies that maximize system performance metrics (e.g., cumulative trip revenue or satisfied user demand). 
Despite their diverse methodologies, most methods share a common practice: scheduling is generally performed only once or twice daily (sometimes even less frequently), usually during off-peak periods with low user activity. This scheduling frequency is driven mainly by budget limitations and regulatory policies—for example, the National Association of City Transportation Officials requires operators to complete fleet rebalancing by 5 a.m. daily to ensure adequate vehicle availability~\cite{NACTO2023micromobility}.
However, this practice introduces a fundamental limitation. User energy preferences and mobility demand can vary significantly and unpredictably across space and time~\cite{heumann2021spatiotemporal,pamminger2024experimental}, causing plans prepared hours in advance to become ineffective in the face of unforeseen events like holidays, local events, or sudden weather changes (see Section~\ref{sec:motivation}). Although some recent methods explicitly incorporate robustness considerations~\cite{tan2023joint,tan2024robust,yan2024robust}, they still struggle to effectively handle large, sudden shifts in demand or to respond swiftly enough.

\textbf{Opportunity and Key Idea}:
Autonomous shared micromobility vehicles (ASMVs), as an emerging form of shared micromobility, possess the capability of self-rebalancing~\cite{sanchez2020autonomous,kondor2021estimating,coretti2023urban,wu2025towards}. 
This capability allows for real-time vehicle redistribution, effectively addressing unexpected demand and significantly enhancing traditional low-frequency redistribution approaches. 
Considering the high cost of deploying ASMVs~\cite{sanchez2020autonomous} and the existing benefits provided by traditional vehicles already in operation, our objective is to integrate a minimal number of ASMVs to complement traditional shared micromobility services. 
It is important to clarify that our approach does not focus on providing on-demand ASMV services like robotaxis, where vehicles directly respond to user requests. 
Instead, we periodically (e.g., hourly) redistribute a small fleet of ASMVs to strategically adjust the overall supply distribution, presenting a more cost-effective solution compared to extensive deployments required for robotaxi-like services.

\textbf{Our Work}: We introduce SMART, a \underline{s}cheduling fra\underline{m}ework for sh\underline{a}red mic\underline{r}omobility services that in\underline{t}egrates ASMVs with conventional vehicles.
The primary objective of this framework is to seamlessly incorporate ASMV operations into existing scheduling practices without altering established strategies. 
Achieving this seamless integration introduces a key challenge: developing an adaptive scheduling approach for ASMVs that respects and complements existing scheduling frameworks. 
This challenge is addressed through two specific questions: (1) how to optimally determine the initial distribution of ASMVs across the city; and (2) how to effectively coordinate their real-time self-rebalancing to enhance overall system performance. 
To resolve these questions, we design a hierarchical reinforcement learning framework consisting of two interconnected levels:
(1) High-level: ASMV redistribution to determine the initial distribution of ASMVs by leveraging current distributions of traditional micromobility vehicles and predicted user demand.
each ASMV acts as an individual agent that learns 
(2) Low-level: ASMV self-rebalancing to rebalance each individual ASMV based on the shared global information, including real-time vehicle distributions and demand forecasts.
Both hierarchical levels share a unified objective function aimed at maximizing the overall service performance of the entire micromobility system.
The key contributions of this work are as follows:

(1) We introduce and explore the concept of integrating autonomous shared micromobility vehicles (ASMVs) with traditional scheduling methods, aiming for seamless operation of a hybrid micromobility system.

(2) Technically, we design a two-level hierarchical reinforcement learning approach to adaptively schedule ASMVs, respecting and complementing existing scheduling frameworks. The high-level module determines the optimal initial distribution of ASMVs before daily operations, while the low-level module manages real-time self-rebalancing throughout daily use.

(3) We perform comprehensive evaluations using real-world e-scooter data from Chicago. These experiments assess how different numbers of ASMVs influence overall system performance, where we show 3\% ASMVs improve the system by 12.72\% on average. 
Our results also demonstrate strong generalization capability of our framework with different traditional scheduling methods and significantly enhances service performance. 

We hope this work will serve as a foundation and encourage further exploration in the emerging research direction of ASMV-augmented scheduling, 
ultimately contributing to more efficient and responsive urban transportation systems.

%% file: 3Motivation.tex
\section{Preliminary and Motivation}
\label{sec:motivation}
In this section, we first introduce the shared micromobility operation data. 
Then, we motivate our work by analyzing the importance of employing autonomous vehicle scheduling in the current shared micromobility system.

\begin{table}[t]
\caption{\small Samples in the dataset}
\vspace{-5pt}
\label{tb:data}
\resizebox{\linewidth}{!}{
\begin{tabular}{cccc}
\toprule
Trip ID          & Start Time       & End Time   & Trip Distance (m)                                                                                                                                  \\ 
T001     & 5/28/2022 14:00  & 5/28/2022 15:00  & 2,484  \\ \midrule
Trip Duration (s)   &  Start Region     & End Region       & Vehicle Operator                                                                                                                         \\ 
1,544 & -87.62519, 41.87887 & -87.62520 41.87886 & Lime  \\
\bottomrule 
\end{tabular}
}
\end{table}
\vspace{-5pt}

\subsection{Data Description}
\label{subsec:data}
In this work, we employ a publicly accessible dataset released by the City of Chicago \cite{chicago_data}, comprising over 629,000 e-scooter trips operated by Lime, Spin, and Bird between June and September 2022. The dataset contains information such as trip time, distance, operator ID, departure time, and region, among other attributes. This information is regularly uploaded by the operators to the city’s Department of Transportation for oversight and policy development.

\subsection{Why Autonomous Vehicle Scheduling?}
Our research is built upon the hypothesis that existing vehicle scheduling models for shared micromobility systems are generally not robust to sudden or unexpected changes in demand, primarily due to their low-frequency execution.
To investigate this hypothesis, we evaluate the performance of various scheduling approaches using real-world trip data from the City of Chicago. Specifically, we measure the effectiveness of each approach through the demand satisfaction rate, defined as the average ratio of satisfied demand to total demand across all regions per day. Because actual user demand is typically unobservable, we follow standard practice from prior studies~\cite{wang2021record,li2021dynamic,tan2023joint} and use recorded trips as a proxy for total demand. Additionally, we examine the impact of incorporating unobserved or background demand on scheduling performance in Section~\ref{subsec:demand}.
We compare three distinct scheduling methods: (1) RECOMMEND~\cite{tan2023joint}, a multi-agent reinforcement learning (MARL) framework tailored for micromobility scheduling; (2) GA, a genetic algorithm-based approach optimizing vehicle deployment using predicted user demand and the total available vehicles; and (3) SDSM, a static demand-supply matching method relying on historical demand data for vehicle rebalancing.

\begin{figure}[h]
\vspace{-10pt}
    \centering
    \includegraphics[width=1.0\linewidth, keepaspectratio=true]{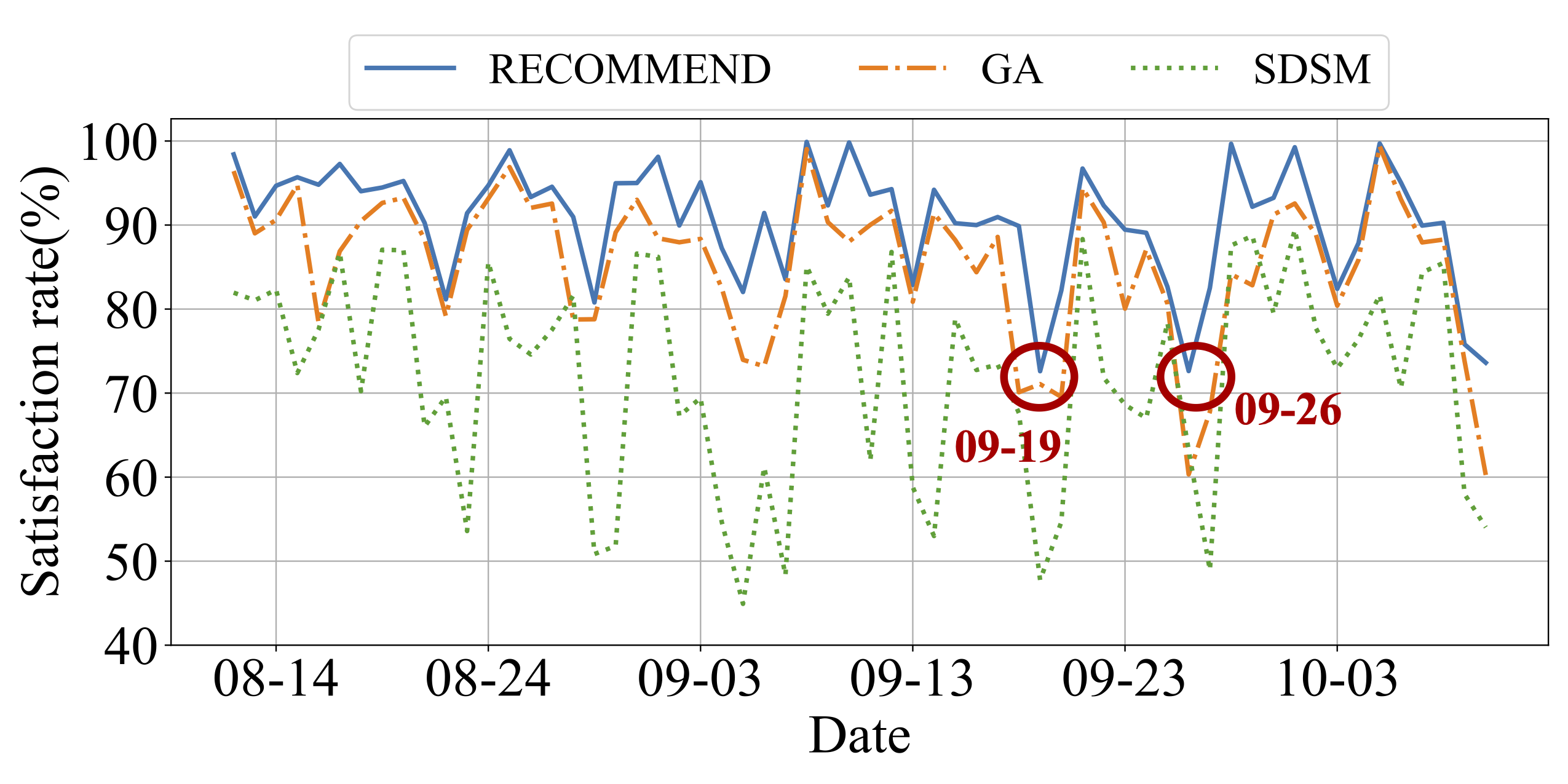}
    \caption{The average demand satisfaction rate of different baselines in two months}
    \label{fig:motivation_p1}
\vspace*{-10pt}
\end{figure}

Figure~\ref{fig:motivation_p1} illustrates the daily demand satisfaction rates achieved by these three methods across all 77 regions over a two-month period (08/11/2022–10/10/2022). Our analysis shows that the optimization-based method (GA) generally outperforms the static scheduling method (SDSM), while the multi-agent reinforcement learning-based method (RECOMMEND) consistently delivers the best performance. The relatively poor performance of SDSM can be attributed to its heavy reliance on historical demand patterns, making it susceptible to unpredictable spatial-temporal fluctuations. The GA method, although optimized based on predicted demand, lacks dynamic modeling of the actual service process, limiting its effectiveness in addressing real-time regional imbalances.

While RECOMMEND achieves a strong overall performance with an average satisfaction rate of 90.61\%, we observe significant performance drops on particular days, such as September 19 and September 26. 
To better understand these drops, we isolate low-performing days and analyze their spatial-temporal trip distributions, aiming to identify how irregular demand patterns affect the robustness of the scheduling policies.

\begin{figure}[h]
    \centering
    \includegraphics[width=1\linewidth, keepaspectratio=true]{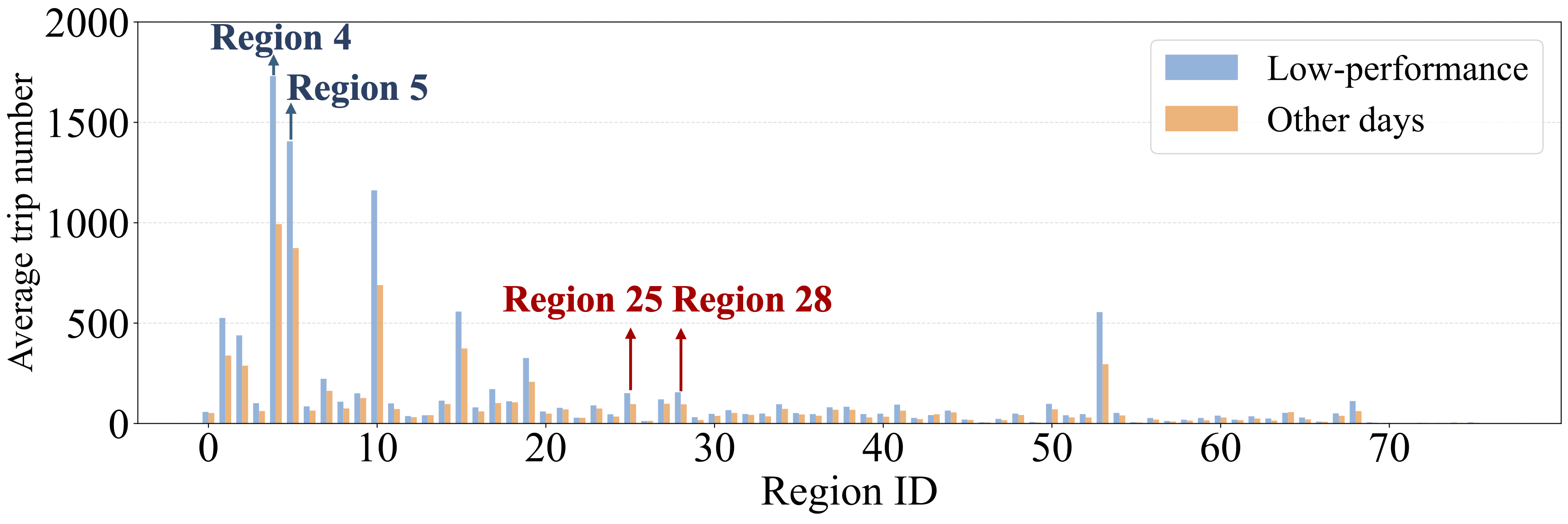}
    \caption{The differences in trip number between low-performance and other days from the spatial perspective}
    \label{fig:moti_spatial}
\vspace*{-10pt}
\end{figure}

\noindent \textbf{Spatial Perspective}: Figure~\ref{fig:moti_spatial} shows the difference in average trip volume across regions between low-performance days and other days. Here, low-performance days are those days when RECOMMEND achieves a demand satisfaction rate lower than 85\%. We observe that some regions experience significantly higher demand on low-performance days, including both typically low-demand regions (e.g., Region 25 and Region 28) and core high-demand regions (e.g., Region 4 and Region 5). This indicates not only a spatial shift in vehicle mobility but also abnormal demand spikes in some high-demand areas. Compared with the trip numbers during other days, there is a total increase of 50.91$\%$ across all regions during low-performance days.
Since the multi-agent reinforcement learning model relies on infrequent vehicle scheduling and learns from historical averages, it may fail to anticipate such irregular surges. As a result, the system suffers from localized vehicle shortages—even in familiar high-demand zones—leading to reduced satisfaction during these low-performing days.

\begin{figure}[h]\centering
\begin{minipage}[h]{0.46\linewidth}
    \includegraphics[width=\linewidth, keepaspectratio=true]{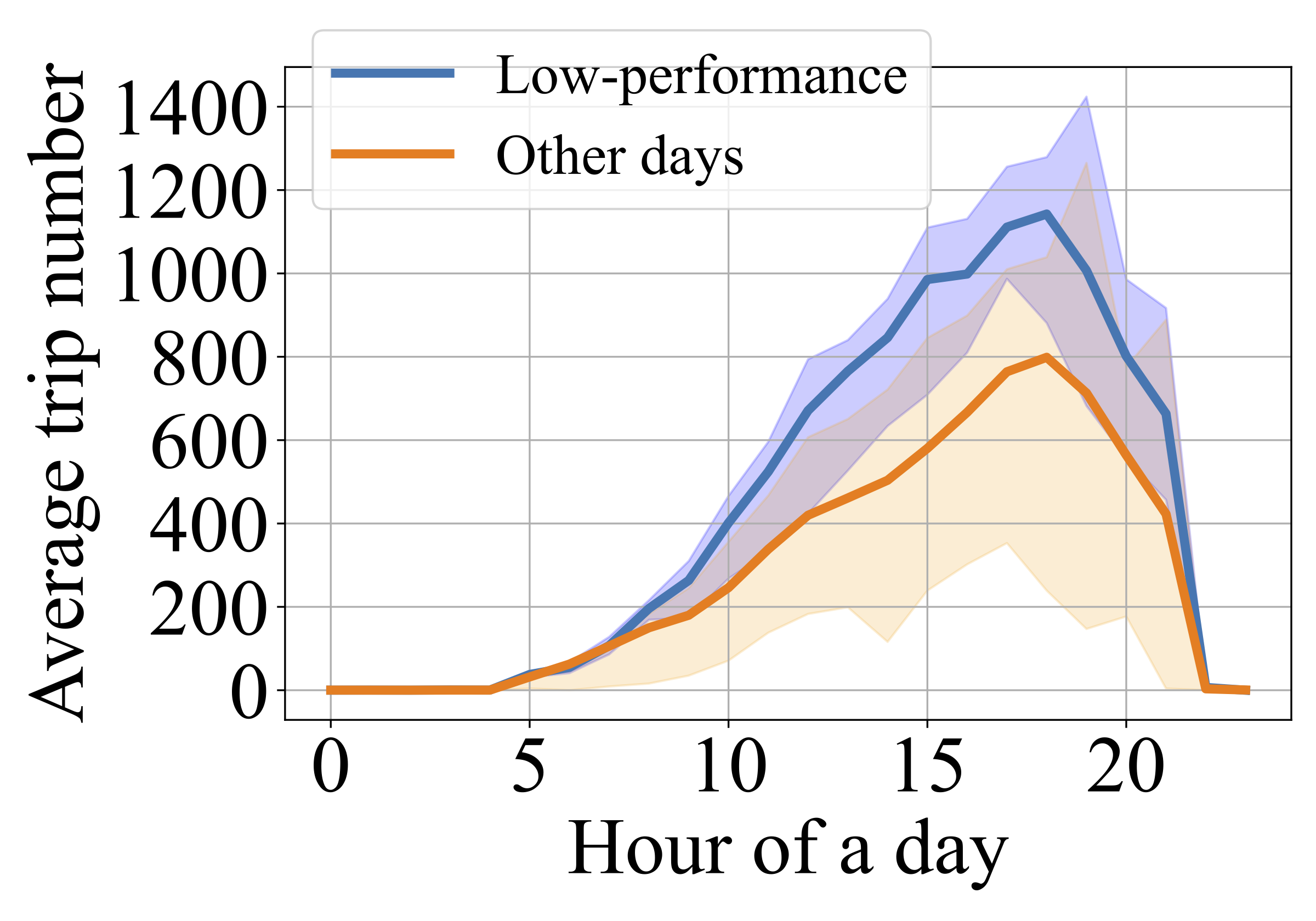}
    \vspace{-15pt}
    \captionsetup{font={small}}
    \caption{The differences in trip number between low-performance days and other days from the temporal perspective}
    \label{fig:moti_temporal}
\end{minipage}
\hspace{10pt}
\begin{minipage}[h]{0.44\linewidth}
    \includegraphics[width=\linewidth, keepaspectratio=true]{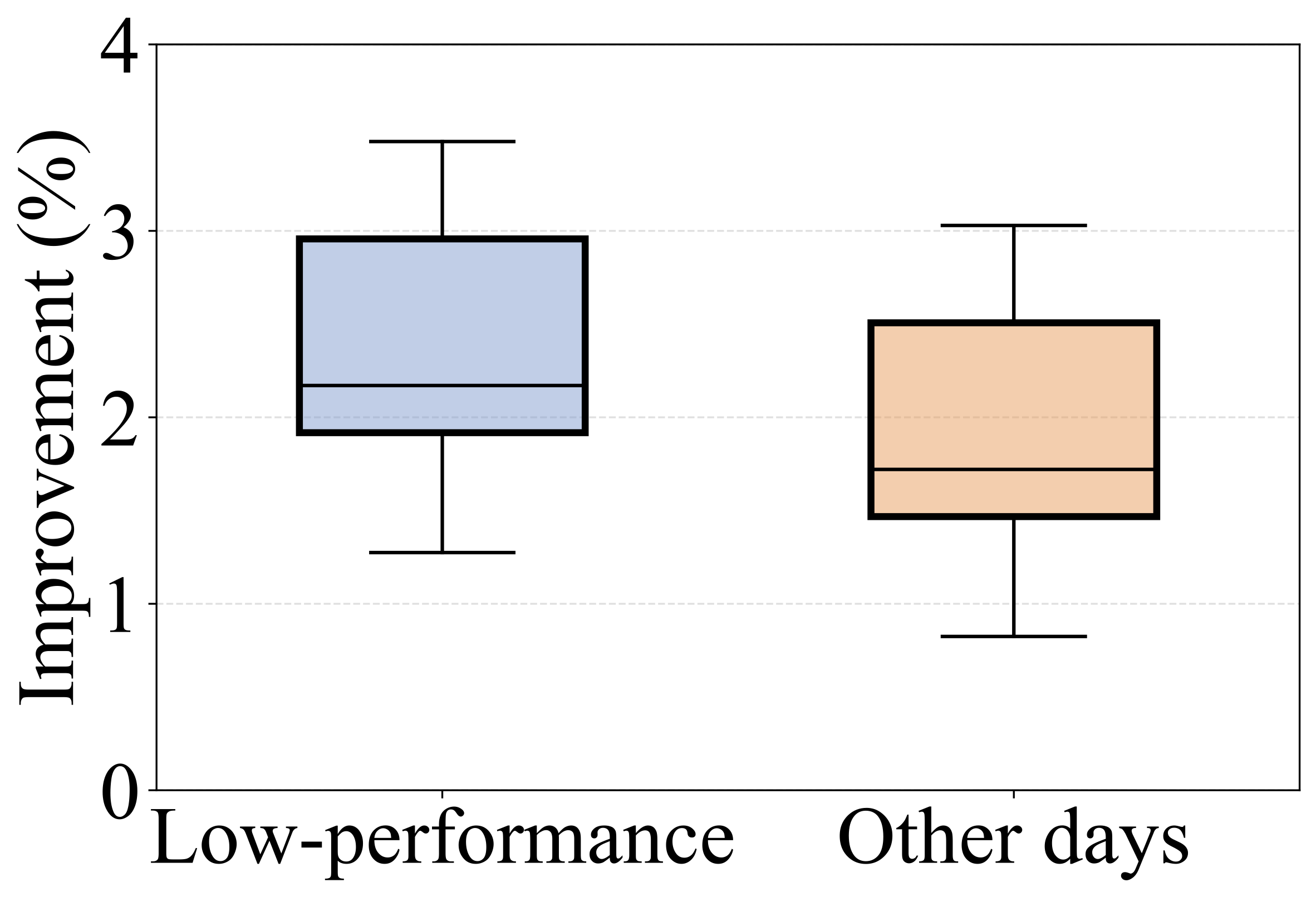}
    \vspace{-15pt}
    \captionsetup{font={small}}
    \caption{The distribution of system performance improvements between low-performance days and other days}
    \label{fig:motivation_p3}
\end{minipage}
\vspace{-5pt}
\end{figure}

\noindent \textbf{Temporal Perspective}: as shown in Figure~\ref{fig:moti_temporal}, low-performance days exhibit more pronounced temporal peaks (i.e., 11 a.m.$\sim$8 p.m.), with trip volumes spiking sharply during specific hours. This indicates that demand is highly concentrated in a few time windows, posing challenges for scheduling strategies to allocate sufficient vehicles to meet such sudden surges. The MARL-based policy, though effective on average, appears less responsive to these high-variance patterns, leading to the lack of supply during peak periods.

These findings highlight a key limitation of the existing scheduling framework: while effective under regular and predictable demand conditions, it struggles to maintain robustness in the face of sharp spatial-temporal demand fluctuations. The infrequent vehicle scheduling makes it unable to anticipate and respond to sudden regional surges, leading to vehicle shortages and reduced system performance on certain days.

\noindent \textbf{Motivation of ASMVs}: 
We explore the use of ASMV scheduling, where vehicles possess self-rebalancing capabilities to relocate themselves based on the real-time vehicle distribution and predicted user demand. To motivate this approach, we implement a simplified autonomous scheduling model and integrate it into the existing shared micromobility system. 
In this design, a portion of vehicles (i.e., each region is assigned an ASMV) are granted the ability to make self-rebalancing decisions throughout the day, allowing them to relocate themselves to the nearby regions where there is no available vehicle. To further evaluate the effectiveness of integrating ASMV scheduling, we compare the system's performance improvement across low-performance days and other days.

\vspace{-5pt}

Figure~\ref{fig:motivation_p3} confirms that incorporating simple autonomous vehicle scheduling can improve overall system performance across both scenarios, with an increase of average satisfaction rate from 90.61$\%$ to 92.57$\%$. The enhancement is particularly significant on low-performance days, demonstrating the effectiveness of ASMV scheduling in mitigating severe demand-supply mismatches. However, the results also reveal that simply reacting to the lack of neighboring supply is not sufficient to fully align the vehicle distribution with dynamic user demand, which motivates the need for a more sophisticated ASMV scheduling method.

%% file: 8ProblemDefinition.tex
\section{Problem Formulation}

\textbf{Problem Setting}: We partition a city into $N$ regions according to the official community divisions~\cite{chicago_data}. A day is divided into $T$ equal-length time intervals. Considering the shared micromobility system operates both traditional and autonomous shared micromobility vehicles, we use $S_{i,t}^{trad}$ and $S_{i,t}^{auto}$ to denote the number of traditional and autonomous vehicles in region $i$ at the beginning of timeslot $t$ for $1\leq i\leq N$, respectively. We use $S_{t}^{trad}\in \mathbb{N}^{N}$ ($S_{t}^{trad}=S_{i,t}^{trad}$,$\forall$ $i\in N$) and $S_{t}^{auto}\in \mathbb{N}^{N}$ ($S_{t}^{auto}=S_{i,t}^{auto}$,$\forall$ $i\in N$) to denote the joint traditional and autonomous vehicle distribution in the city, respectively.  We assume that both types of vehicles jointly serve user demand in the city. Therefore, we use $U_{t}^{i,j}$ ($1\leq i,j\leq N$) to denote user demand, which quantifies the number of user requests from region $i$ to region $j$. 

\textbf{Scheduling}: To meet the highest possible number of future user requests, the system operator rebalances vehicles prior to daily operations. Specifically, at the beginning of each day (i.e., before the first time interval), the operator determines the initial distributions of traditional denoted by $S_{0}^{trad}$, based on the current traditional vehicle distribution and future user demand prediction. Since autonomous vehicles are assumed to be introduced to support the existing system, their deployment $S_{0}^{auto}$ should take the traditional vehicle deployment $S_{0}^{trad}$ and future user demand into consideration. Their formulations are as follow:
\begin{equation}
    S_{0}^{trad} \leftarrow f_{reb}^{trad}(S_{pre}^{trad}, U_{1:T}), \quad  S_{0}^{auto} \leftarrow f_{reb}^{auto}(S_{0}^{trad},U_{1:T}),
\end{equation}
where $S_{pre}^{trad}$ represents the traditional vehicle distribution before the daily system operations. $f_{reb}^{trad}$ and $f_{reb}^{auto}$ denote the rebalancing strategies for traditional and autonomous vehicles, respectively. During the daily system operations, at the beginning of each time interval, each autonomous vehicle $k$ rebalances itself according to its current location, the global vehicle distributions, and predicted demand in the next interval, formulated as:
\begin{equation}
    z_{t}^{k+1} \leftarrow f^{auto}_{k}(z_{t}^{k},S_{t}^{trad},S_{t}^{auto},U_{t:t+1}),
\end{equation}
where $z_{t}^{k}$ represents the location of the autonomous vehicle $k$ at the beginning of time inteval $t$. $f^{auto}_{k}$ denotes the self-rebalancing strategy of the autonomous vehicle $k$. It is worth noting that traditional vehicles follow a static scheduling scheme (e.g., rule-based and optimization-based), where the deployment determined at the beginning of the day remains unchanged during the subsequent operations. The autonomous vehicle distribution $S_{t}^{auto}$ is aggregated from individual vehicle locations $\{z_t^k\}_{k=1}^{K}$:
\begin{equation}
    S_{t}^{auto} = \text{Aggregate}(\{z_t^k\}_{k=1}^{K}),
\end{equation}
where $K$ is the total number of autonomous vehicles.

\textbf{System Operation}: During daily system operation, users continuously request trips from one region to another, and only a portion of user demands can be satisfied due to vehicle availability constraints. We denote $D_t$ as the total number of satisfied user demand across all regions in the time interval $t$:
\begin{equation}
    D_{t} = f_{trip}(S_{t}^{trad},S_{t}^{auto},U_{t}),
\end{equation}

\textbf{Objective}: In our shared micromobility system, the traditional vehicle scheduling strategy $f_{reb}^{trad}$ is assumed to be pre-determined by existing works~\cite{tan2023joint,lam2016autonomous}. Our goal is to develop an optimal algorithm to efficiently determine the initial deployment of autonomous vehicles and individual rebalancing strategies, to maximize the cumulative user demand satisfaction rate $D_{rate}$. We define the objective function as follows:
\begin{equation}
\label{eq:objective}
    argmax_{f_{reb}^{auto}, \{f_k^{auto}\}_{k=1}^K} \; D_{rate} =  \frac{\sum_{t=1}^{T} f_{trip}(S_{t}^{trad},S_{t}^{auto},U_{t})}{\sum_{t=1}^{T} U_{t}}.
\end{equation}

%% file: 4Method.tex
\section{Methodology}

\begin{figure}[h]
    \centering
    \includegraphics[width=0.9\linewidth, keepaspectratio=true]{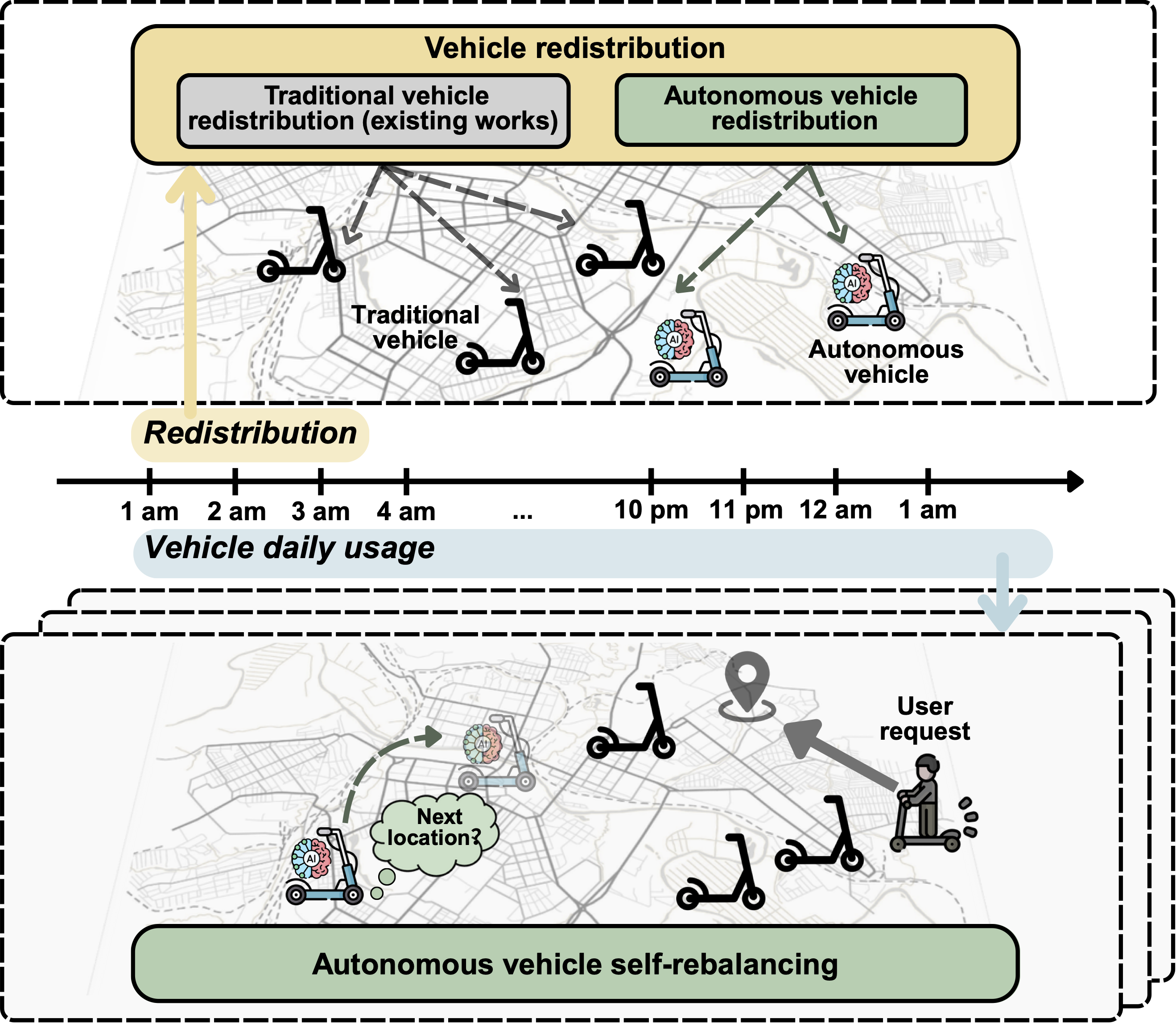}
    \vspace{-5pt}
    \caption{Overview of hierarchical RL framework for rebalancing autonomous shared micromobility vehicles}
    \label{fig:framework}
    \vspace{-15pt}
\end{figure}

\subsection{Design Overview}

We design a two-level hierarchical RL-based framework for rebalancing autonomous shared micromobility vehicles, as shown in Figure~\ref{fig:framework}. This framework consists of three components: the environment, vehicle redistribution, and autonomous vehicle self-rebalancing. (1) \textbf{Environment}: It simulates the shared micromobility system operations, including vehicle dynamics, user demand, and request fulfillment. (2) \textbf{Vehicle redistribution}: It is responsible for the redistribution of shared micromobility vehicles across the entire city. This component consists of traditional and autonomous vehicle redistribution. (i) In our work, we regard traditional vehicle redistribution as a modular component that follows existing redistribution strategies (e.g., rule-based~\cite{jin2023vehicle,kim2023optimal} and optimization-based~\cite{tan2023joint,li2021dynamic}). We do not alter this module for seamless integration. (ii) We employ a neural network-based agent for the system operator to determine their initial distribution by considering the traditional vehicle distribution and future user demand. (3) \textbf{Autonomous vehicle self-rebalancing}: after the vehicle redistribution, each ASMV is assigned an agent to make sequential decisions about its next location at the beginning of each time interval, based on its current location, the global vehicle distributions, and predicted demand in the next time interval. All individual agents share a common policy network. Both autonomous vehicle redistribution and vehicle self-rebalancing agents are trained with a shared objective: maximizing the cumulative demand satisfaction rate.

\subsection{Two-level Hierarchical RL-based vehicle scheduling}

We formulate the autonomous vehicle scheduling problem as a two-level hierarchical Markov Decision Process, denoted as $\mathcal{M}^{high}$$=\{\mathcal{S}^{high}, \mathcal{A}^{high}, \mathcal{R}, \mathcal{P}^{high}, \gamma^{high}\}$ and $\mathcal{M}^{low}$$=\{\mathcal{S}^{low}, \mathcal{A}^{low},$ $ \mathcal{R}, \mathcal{P}^{low},$ $ \gamma^{low}\}$, to model the high-level autonomous vehicle redistribution and low-level vehicle self-rebalancing, respectively. Both levels share a common objective: to maximize the cumulative demand satisfaction rate. The definitions of these notations are as follows:

\noindent \textbf{1. High-level MDP}
\begin{itemize}
    \item \textbf{Agent}: We employ a centralized agent to make decisions about how to redistribute autonomous shared micromobility vehicles across all the regions.
    \item \textbf{State $\mathcal{S}^{high}$}: At the beginning of each day, the high-level agent observes its own state $s^{high}_{0}$, containing the scheduled traditional vehicle distribution $S_{0}^{trad}$ and predicted future h-timeslot user demand $U_{1:h}$ before the next autonomous vehicle redistribution. The future demand is predicted by a pre-trained prediction model~\cite{tan2023joint}.
    \item \textbf{Action $\mathcal{A}^{high}$}: The high-level agent takes the action $a^{high}_{0}$ of the initial redistribution decision for autonomous vehicles across $N$ regions, that is, the number of vehicles to be allocated in each region at the beginning of each day. We assume that the redistribution strategies of both traditional and autonomous shared micromobility vehicles are executed by trucks that move vehicles between regions.
    \item \textbf{Transition $\mathcal{P}^{high}$}: It is deterministic, where the agent's action defines the initial distribution $S_{0}^{auto}$ for each day.

\end{itemize}

\noindent \textbf{2. Low-level MDP}
\begin{itemize}
    \item \textbf{Agent}: Each autonomous vehicle is assigned an agent to make decisions about how to rebalance itself during the daily system operations. Instead of utilizing a single agent to control the rebalancing strategies for the entire city, We utilize centralized training and decentralized execution to reduce the computational complexity~\cite{lowe2017multi,he2023robust}.
    \item \textbf{State $\mathcal{S}^{low}$}: At the begining of the time interval $t$, the state of the vehicle agent $i$ is defined as $s^{low}_{i,t}=\{z_{t}^{i},S_{t}^{trad},S_{t}^{auto},U_{t:t+1}\}$, where $z_{t}^{i}$ denotes the spatial location of vehicle $i$ at the begining of the time interval $t$. $S_{t}^{trad}$ and $S_{t}^{auto}$ denote the traditional and autonomous vehicle distributions at the beginning of the time interval $t$, respectively. $U_{t:t+1}$ denotes the predicted demand in the time interval $t$.
    \item \textbf{Action $\mathcal{A}^{low}$}: Given the above state, the vehicle agent $i$ at the beginning of the time interval $t$ decides the rebalanced location before the next time interval $t+1$, denoted as $a^{low}_{i,t}$. The process of self-rebalancing in autonomous vehicles incurs an intrinsic energy cost. We assume that if the remaining energy of an autonomous vehicle falls below a specified threshold, it is considered unavailable for subsequent self-rebalancing and daily use until recharged in the next autonomous vehicle deployment cycle.
    \item \textbf{Transition $\mathcal{P}^{low}$}: It denotes the probability that the joint state $s^{low}_{t}$ $(s^{low}_{t}=\{s^{low}_{i,t}\}_i)$ transfers to the next joint state $s^{low}_{t+1}$ given the joint action $a^{low}_t$ $(a^{low}_{t}=\{a^{low}_{i,t}\}_i)$.
\end{itemize}

\textbf{Reward $\mathcal{R}$}: In our work, we desire to make two-level agents cooperatively maximize the shared micromobility service performance. Therefore, the two-level agents share the same reward function, that is, the cumulative demand satisfaction rate:
\begin{equation}
R_{t} 
\begin{cases}
\frac{D_{1:T}}{U_{1:T}}, & \text{if } t = T \\
0, & \text{otherwise}
\end{cases}
\end{equation}
where $D_{1:T}$ and $U_{1:T}$ denote the number of satisfied demand and total demand from time interval $1$ to time interval $T$, respectively. Due to the differences in task granularity and decision frequency, the high-level agent receives an immediate reward after each deployment decision, whereas low-level individual agents only receive a sparse terminal reward at the end of the day.

\textbf{Discounted Factor $\gamma$}: The discount factor $\gamma \in [0,1)$ determines how much the agent values future rewards relative to immediate ones. When $\gamma=0$, the agent focuses solely on immediate rewards, favoring actions that yield instant gains. In contrast, as $\gamma$ approaches 1, the agent places increasing emphasis on long-term outcomes, learning strategies that maximize cumulative rewards over time.

\textbf{Objective}: We aim to optimize a two-level hierarchical reinforcement learning (HRL) framework to maximize the overall user demand satisfaction rate. The framework consists of (i) a high-level deployment planning agent and (ii) low-level individual rebalancing agents. (i) The goal of the high-level agent is to learn a policy $\pi_H(S_0^{auto} \mid s_0^{high})$ that determines the initial distribution of autonomous vehicles before daily operations, to maximize the expected cumulative return: $J_H(\pi_H) = \mathbb{E}_{\pi_H} \left[ \sum_{t=0}^{T} \gamma^t R_t \right] = \mathbb{E}_{\pi_H} [R]$. To solve this high-level deployment planning task, we define the value functions, including the state-value function $V^{\pi_{H}}(s^{high}_0)$ and the state-action value function (Q-function) $Q^{\pi_H}(s^{high}_0, a^{high}_0)$:
\begin{equation}
    V^{\pi_{H}}(s^{high}_0)=\mathbb{E}_{\pi_H} [\sum_{t=0}^{T}\gamma^{t}R_{t} \mid s_0^{high}],
\end{equation}

\begin{equation}
    Q^{\pi_H}(s^{high}_0, a^{high}_0) = \mathbb{E}_{\pi_H} [\sum_{t=0}^{T}\gamma^{t} R_{t} \mid s^{high}_0, a^{high}_0].
\end{equation}

(ii) Each low-level agent $i$ makes sequential decisions over the time horizon of an episode. All agents share the same policy $\pi_L$ and aim to maximize the global reward shared with the high-level deployment planning agent $R$: $J_L(\pi_L) = \mathbb{E}_{\pi_L} \left[ \sum_{t=1}^{T} \gamma^t R_t \right] = \mathbb{E}_{\pi_L} [R]$. We define the low-level value functions, including the state-value function $V^{\pi_L}(s_{i,t}^{low})$ and the state-action value function (Q-function) $Q^{\pi_L}(s_{i,t}^{low}, a_{i,t}^{low})$:

\begin{equation}
    V^{\pi_L}(s_{i,t}^{low}) = \mathbb{E}_{\pi_L} \left[ \sum_{t'=t}^{T} \gamma^{t'-t} R_{t'} \mid s_{i,t}^{low} \right],
\end{equation}
\begin{equation}
    Q^{\pi_L}(s_{i,t}^{low}, a_{i,t}^{low}) = \mathbb{E}_{\pi_L} \left[ \sum_{t'=t}^{T} \gamma^{t'-t} R_{t'} \mid s_{i,t}^{low}, a_{i,t}^{low} \right].
\end{equation}

\subsection{Training Method}

\par The training of our hierarchical reinforcement learning (HRL)-based autonomous vehicle scheduling framework follows a two-level policy structure optimized using the Proximal Policy Optimization (PPO) algorithm~\cite{schulman2017proximal}. Specifically, we decompose the learning problem into a high-level vehicle deployment policy and a low-level autonomous vehicle scheduling policy, each paired with a corresponding value function (critic). To ensure stable and decoupled learning, we apply an alternating update scheme motivated by the existing Nash Equilibrium work \cite{pinto2017robust}. We divide each iteration into four tasks, as shown in Algorithm~\ref{alg:hrl_ppo}:

\noindent \textbf{Task 1: High-level interaction}: At the start of each episode, before system daily operations, the environment is initialized by sampling a high-level state $s_0^{high} \sim p$, representing the initial traditional vehicle distribution $S_{pre}^{trad}$ (from a modular tradition vehicle scheduling model), and the predicted user demand $U_{1:T}$. The high-level policy samples an action to be executed in the environment $a_{0}^{high}$, determining the initial deployment of autonomous vehicles.

\noindent \textbf{Task 2: Low-level interaction}: After deployment, the system enters the low-level control phase. During system daily operations, At the beginning of each time interval $t=1,...,T$, each autonomous vehicle $k=1,...,K$ observes its local state $s_{k,t}^{low}$, composed of its current location $z_{t}^{k}$, the system-wide vehicle distribution ($S_{t}^{trad}$, $S_{t}^{auto}$), and predicted user demand before the next time interval $U_{t:t+1}$. Each vehicle samples and execute an action $a^{low}_{k,t} \sim \pi_L(\cdot | s^{low}_{k,t})$ (i.e., the next location to move). The environment then transitions to the next state, and we store all the trajectories from the high-level agent and low-level agents into the replay buffer $\mathcal{D}$.

\noindent \textbf{Task 3: Policy update}: Based on a predefined update frequency $N_{\text{freq}}$, we adopt an alternating training scheme to update the two-level policies. 
(1) For the high-level policy, we compute the Monte Carlo advantage~\cite{mnih2016asynchronous} as 
$\hat{A}_0^{\text{high}} = R_T - V_H(s_0^{\text{high}})$, 
which is suitable given that the high-level agent interacts with the environment only once per episode and its action affects the long-term return. 
(2) For the low-level policy, we employ Generalized Advantage Estimation (GAE)~\cite{schulman2015high} to compute smoother and lower-variance advantages based on temporal difference (TD) errors:
$\delta_t^k = R_t + \gamma V_L(s^{\text{low}}_{k,t+1}) - V_L(s^{\text{low}}_{k,t})$ and
$\hat{A}_t^k = \sum_{l=0}^{T-t-1} (\gamma \lambda)^l \delta_{t+l}^k$.
Using the collected trajectories and estimated advantages, we apply PPO to update the actor-critic networks for each level accordingly.

\begin{algorithm}[h]
\caption{Two-level Hierarchical RL-based Autonomous Vehicle Scheduling with Alternating PPO}
\label{alg:hrl_ppo}
\begin{algorithmic}[1]
\Require Environment $\mathcal{E}$, high-level policy $\pi_H$, low-level policy $\pi_L$, high-level critic $V_H$, low-level critic $V_L$, episode count $N_{ep}$, vehicle count $K$, initial state distribution $p$, learning rate $\alpha$, PPO clip ratio $\epsilon$, discount factor $\gamma$, GAE parameter $\lambda$
\Ensure Initialized policies $\pi_H$, $\pi_L$ and critic networks $V_H$, $V_L$
\For{each episode $e = 1$ to $N_{ep}$}
    \State Receive the initial state $s_0^{high} \sim p$
    \State \textbf{/* Task 1: High-Level Interaction */}
    \State Observe high-level state $s_0^{high}$ ($S_{pre}^{trad}$, $U_{1:T}$)
    \State Sample deployment action $a_0^{high} \sim \pi_H(\cdot | s_0^{high})$
    \State Get value estimate $V_H(s_0^{high})$
    \State Deploy vehicles according to $a_0^{high}$ in $\mathcal{E}$

    \State \textbf{/* Task 2: Low-Level Interaction */}
    \State Initialize agent action list $\mathcal{L}$
    \For{each time step $t = 1$ to $T$}
        \For{each autonomous vehicle $k = 1$ to $K$}
            \State Observe local state $s^{low}_{k,t}$ ($z_t^k$, $S_t^{trad}$, $S_t^{auto}$, $U_{t:t+1}$)
            \State Sample action $a^{low}_{k,t} \sim \pi_L(\cdot | s^{low}_{k,t})$
            \State Get value estimate $V_L(s^{low}_{k,t})$
            \State Add $a_{k,t}^{low}$ to list $\mathcal{L}$
        \EndFor
        \State Execute $\{a_{k,t}^{low}\}_{k=1}^{K}$ in $\mathcal{E}$
        \State Transition to next state, update $S_{t+1}^{auto}$
        \State Store $\{(s^{low}_{k,t}, a^{low}_{k,t}, V_L(s^{low}_{k,t}))\}_{k=1}^{K}$ and $R_t$ into buffer $\mathcal{D}$
    \EndFor
    \State Store high-level tuple $(s_0^{high}, a_0^{high}, R_T, V_H(s_0^{high}))$ into buffer $\mathcal{D}$
    \State \textbf{/* Task 3: Policy Update */}
    \If{$e$ $\bmod$ $N_{freq}$ == 0}
        \State \textbf{// High-Level PPO Update}
        \State Sample minibatch from $\mathcal{D}$: $(s_0^{high}, a_0^{high}, R_T, V_H(s_0^{high}))$
        \State Compute high-level advantage: $\hat{A}_0^{high} = R_T - V_H(s_0^{high})$
        \State Update $\pi_H$ and $V_H$ using PPO with $(s_0^{high}, a_0^{high}, \hat{A}_0^{high}, R_T, \epsilon, \alpha)$
    \Else
        \State \textbf{// Low-Level PPO Update}
        \State Sample minibatch from $\mathcal{D}$: $(s_{k,t}^{low}, a_{k,t}^{low}, V_L(s_{k,t}^{low}), R_t)$
        \State Compute GAE advantage:
        \State $\delta_t^k = R_t + \gamma V_L(s^{low}_{k,t+1}) - V_L(s^{low}_{k,t})$
        \State $\hat{A}_t^k = \sum_{l=0}^{T-t-1} (\gamma \lambda)^l \delta_{t+l}^k$
        \State Update shared $\pi_L$ and $V_L$ using PPO with $(s^{low}_{k,t}, a^{low}_{k,t}, \hat{A}_t^k, R_t, \epsilon, \alpha)$
    \EndIf
\EndFor
\State \Return $\pi_H, \pi_L, V_H, V_L$
\end{algorithmic}
\end{algorithm}

%% file: 5Evaluation.tex
\section{Evaluation}

\subsection{Experiment Setup}
\subsubsection{\textbf{Implementation}} We conduct our experiments on a publicly available real-world shared e-scooter dataset from Chicago~\cite{chicago_data}, which includes data from three operators: Spin, Bird, and Lime. The dataset is divided into two parts: the first two months are used as the training set, and the remaining data are used as the test set. The entire city is partitioned into 77 regions based on the existing community divisions in the dataset. Each day is divided into 24 hours, with traditional shared micromobility vehicle redistribution occurring once per day and autonomous shared micromobility vehicle self-rebalancing taking place once per hour.
In our experiments, we consider three operators (Lime, Spin, and Bird), with 2,695, 2,581, and 2,795 vehicles, respectively. Due to the lack of vehicle remaining energy information in the dataset, we assume a maximum cruising range of 15 miles for shared e-scooters~\cite{tan2025realism}, and we use the trip distance to estimate energy consumption for each recorded trip. We assume that both traditional and autonomous vehicles share the same battery capacity. The charging cost is set at \$4 per e-scooter, accounting for both electricity and labor costs~\cite{FinancialPantherBirdCharger}.
For vehicle usage, only nearby vehicles with enough remaining energy can satisfy the user demand, and the trip revenue is calculated at \$1.00 to unlock + \$0.39 per minute, excluding tax~\cite{chicago_price}. For ASMVs, rebalancing costs are considered as a kind of energy consumption because of their self-rebalancing. traditional vehicles, rebalancing costs are computed based on the existing works~\cite{tan2023joint,tan2024robust}.

We implement our method and baselines with PyTorch 1.9.1, Python-mip 1.14.2, and gym 0.21.0 in a Python 3.7 environment, and we train them on a server equipped with 32 GB of memory and a GeForce RTX 3080 Ti GPU. By testing the performances under different hyperparameter settings, we use the following settings: For the autonomous vehicle redistribution agent, we use an Adam optimizer with an optimal learning rate of 3e-4 among [3e-3, 3e-4, 3e-5]. For the autonomous vehicle self-rebalancing agent, we use an Adam optimizer with an optimal learning rate of 1e-4 among [1e-3, 1e-4, 1e-5]. The clip ratio $\epsilon$ in the PPO update process is set as 0.2. The decay $\lambda$ in GAE computation is set as 0.95. The discount factor $\gamma$ is set as 0.99. The minibatch size is 64 for experience replay. Hyperparameters for other baselines are fine-tuned based on the range in the original papers.

\subsubsection{\textbf{Baselines}} The vehicle scheduling baselines for traditional shared micromobility vehicles are as follows. These baselines are pre-trained and used for traditional vehicle scheduling. 
\begin{itemize}
    \item \textbf{SDSM}. It is a rule-based demand-supply matching method where each operator rebalances the vehicles to each region based on the ratio of the historical demand.
    \item \textbf{GA}~\cite{lam2016autonomous}. It is an optimization-based method that utilizes a genetic algorithm to find the optimal vehicle scheduling strategies to maximize the demand satisfaction rate.
    \item \textbf{RECOMMEND}~\cite{tan2023joint}. It is a state-of-the-art shared electric micromobility vehicle scheduling algorithm that utilizes a MARL-based method to learn the optimal vehicle scheduling polices to maximize the system service performance.
\end{itemize}

We compare our method with the following autonomous vehicle scheduling baselines and variants of our model:
\begin{itemize}
    \item \textbf{IAVS}. It is a rule-based ASMV scheduling method that enables vehicles to relocate themselves, at the beginning of each time interval, to the nearby regions where there is no available vehicle to use.
    \item \textbf{MIP} ~\cite{zhang2016model}. It is a spatial-temporal mixed integer programming method that tries to find the optimal rebalancing strategies to maximize the demand satisfaction rate based on the vehicle distribution and predicted future demand.
\end{itemize}

Variants of our model are as follows:
\begin{itemize}
    \item \textbf{SMART without vehicle redistribution (w/o VRD)}. In this setting, we remove the module of vehicle redistribution and assume the average distribution of ASMVs in the city at the beginning of each episode.
    \item \textbf{SMART without vehicle self-rebalancing (w/o VSR)}. In this setting, we remove the module of vehicle self-rebalancing, and the operator only makes decisions about the redistribution of both autonomous and traditional vehicles at the beginning of each episode.
\end{itemize}

\subsubsection{\textbf{Metric}} The evaluation metric is as follows:
\begin{itemize}
    \item \textbf{Average satisfaction rate}. The average satisfaction rate represents the average ratio of satisfied demand to the total user demand among all the regions (Equation~\ref{eq:objective}).
\end{itemize}

\subsection{Overall Performance}

Table~\ref{tab:overall} shows the overall performance of different combinations of autonomous and traditional vehicle scheduling models. In our experiments evaluating overall system performance, we replace 3$\%$ of traditional shared micromobility vehicles with ASMVs. This proportion is selected in light of the marginal efficiency that these ASMVs contribute to enhancing the performance of the entire shared micromobility system (detailed discussion in Section~\ref{sec:ratio}). Through the experiments, we have the following findings: \textbf{(1) Overall, our model SMART achieves the most significant performance improvement over all three traditional vehicle scheduling baselines}, achieving an improvement of at least 7.56$\%$ in demand satisfaction rate. The superior performance of SMART is attributed to its two-level design: the module of ASMV redistribution optimizes the initial allocation of a limited number of ASMVs based on the current traditional vehicle distribution and the predicted future demand. Compared to IAVS, a rule-based method that reacts to immediate shortages in a myopic manner, this anticipatory deployment allows ASMVs to be strategically positioned in areas with potential supply shortages, thereby facilitating more effective self-rebalancing in subsequent time intervals. Furthermore, unlike MIP, which relies solely on centralized optimization, Furthermore, unlike MIP, which relies solely on centralized optimization, SMART incorporates a decentralized self-rebalancing policy at the vehicle level, enabling fine-grained and context-aware relocation decisions. As a result, the system is better prepared to respond to spatial-temporal demand fluctuations throughout the day.
\textbf{(2) SMART shows greater marginal gains when paired with weaker traditional vehicle scheduling baselines}. For instance, under SDSM—where the initial supply-demand mismatch is more severe—the relative improvement of SMART over SDSM without ASMVs introduced reaches 30.30$\%$, whereas under RECOMMEND the improvement is 7.56$\%$, which highlights SMART’s capacity to correct substantial supply gaps caused by suboptimal traditional vehicle scheduling, validating its robustness and scalability across different deployment qualities.
However, despite the substantial relative improvements, the system-level performance under weaker baselines (e.g., SDSM) remains lower than that achieved under stronger baselines (e.g., RECOMMEND), with SMART only reaching 89.84$\%$ under SDSM. This is primarily due to the limited number of ASMVs, which constrains the extent to which autonomous vehicle scheduling can compensate for severe supply shortages. Nevertheless, the consistently strong performance improvement, even under such a constrained setting, suggests that even a small-scale deployment of ASMVs, when effectively coordinated, can yield significant system-wide benefits.

\begin{table}[t]
\caption{Performance comparison (i.e., demand satisfaction rate) of different combinations of autonomous and traditional vehicle scheduling models. - indicates no ASMVs used.}
\label{tab:overall}
\vspace{-5pt}
\resizebox{\linewidth}{!}{
\begin{tabular}{lccc}
\toprule
\multirow{2}{*}{\shortstack{\textbf{ASMV} \\ \textbf{Scheduling}}} & \multicolumn{3}{c}{\textbf{Traditional Vehicle Scheduling Models}} \\
 & \textbf{SDSM} & \textbf{GA} & \textbf{RECOMMEND} \\
\midrule
-        & 68.95$\%$ ($\pm$0.35) & 82.05$\%$ ($\pm$0.36) & 90.61$\%$ ($\pm$0.41)  \\ 
IAVS     & 80.09$\%$ ($\pm$0.40) & 86.78$\%$ ($\pm$0.39) & 92.57$\%$ ($\pm$0.43) \\ 
MIP      & 83.68$\%$ ($\pm$0.38) & 87.74$\%$ ($\pm$0.37) & 93.26$\%$ ($\pm$0.42) \\ 
w/o VRD  & 85.82$\%$ ($\pm$0.41) & 90.21$\%$ ($\pm$0.44) & 95.84$\%$ ($\pm$0.46) \\ 
w/o VSR  & 80.20$\%$ ($\pm$0.38) & 86.99$\%$ ($\pm$0.41) & 90.87$\%$ ($\pm$0.44) \\ \midrule
\textbf{SMART} & \textbf{89.84$\%$ ($\pm$0.44)} & \textbf{92.39$\%$ ($\pm$0.48)} & \textbf{97.46$\%$ ($\pm$0.50)}  \\ 
\bottomrule
\end{tabular}
}
\end{table}

\begin{figure*}[t]
  \centering

  \begin{minipage}[t]{0.24\textwidth}
    \centering
    \includegraphics[width=\linewidth]{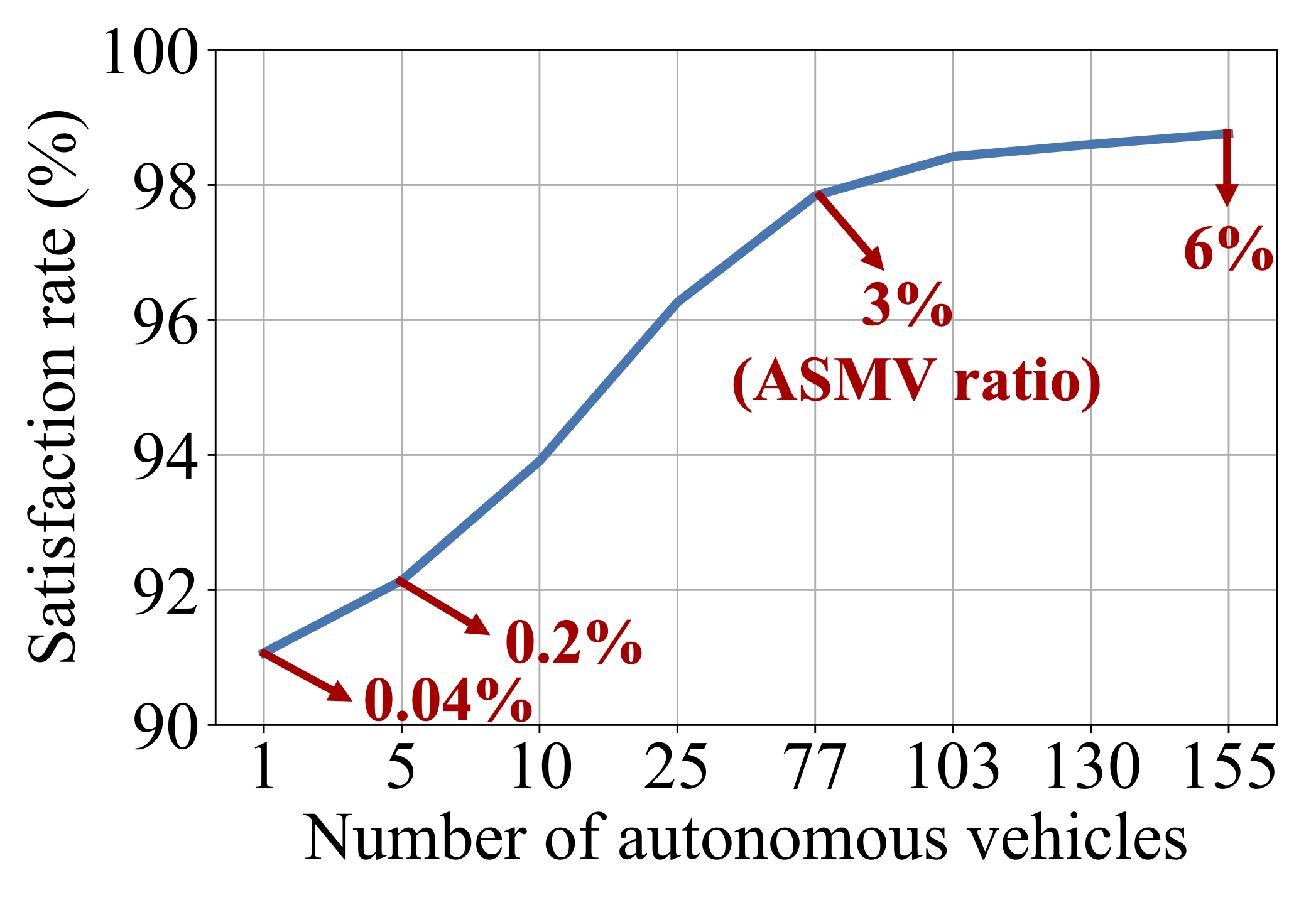}
    \vspace{-10pt}
    \caption{The effect of different scales of ASMVs}
    \label{fig:ratio}
  \end{minipage}
  \hfill
  \begin{minipage}[t]{0.24\textwidth}
    \centering
    \includegraphics[width=\linewidth]{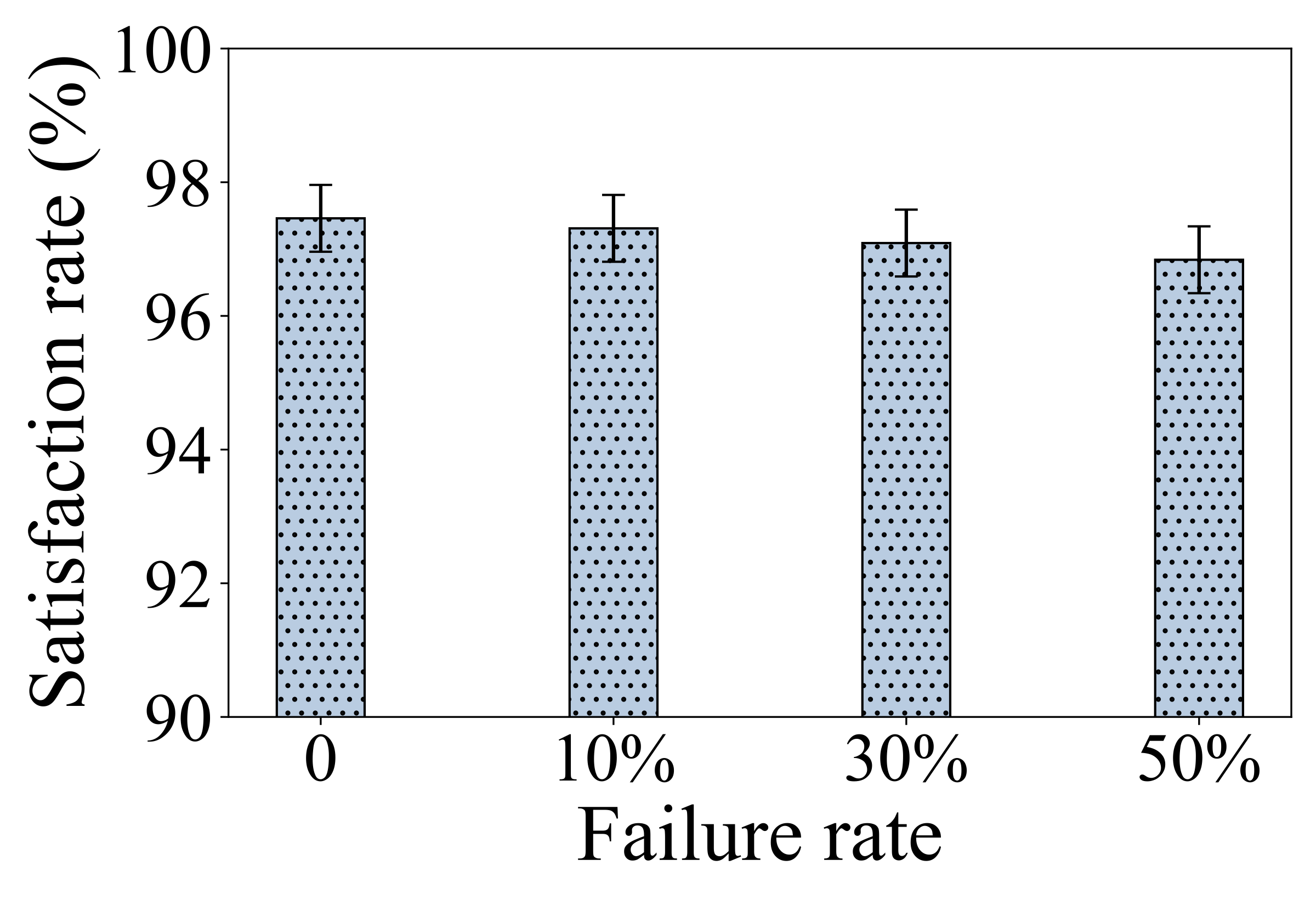}
    \vspace{-10pt}
    
    \caption{The effect of different failure rates of ASMVs}
    \label{fig:fault}
  \end{minipage}
  \hfill
  \begin{minipage}[t]{0.24\textwidth}
    \centering
    \includegraphics[width=\linewidth]{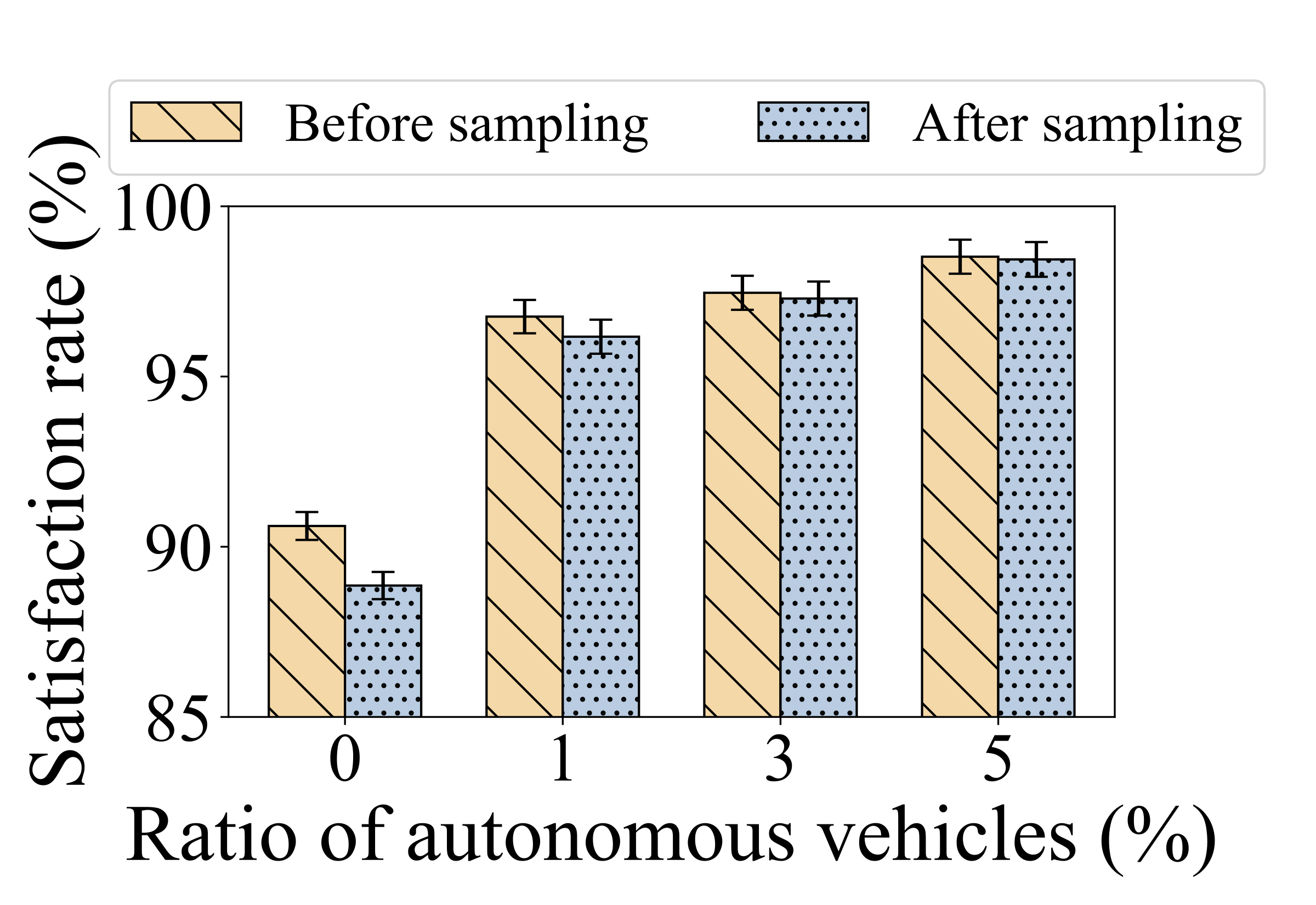}
    \vspace{-10pt}
    
    \caption{The effect of background demand sampling}
    \label{fig:sampling}
  \end{minipage}
  \hfill
  \begin{minipage}[t]{0.24\textwidth}
    \centering
    \includegraphics[width=\linewidth]{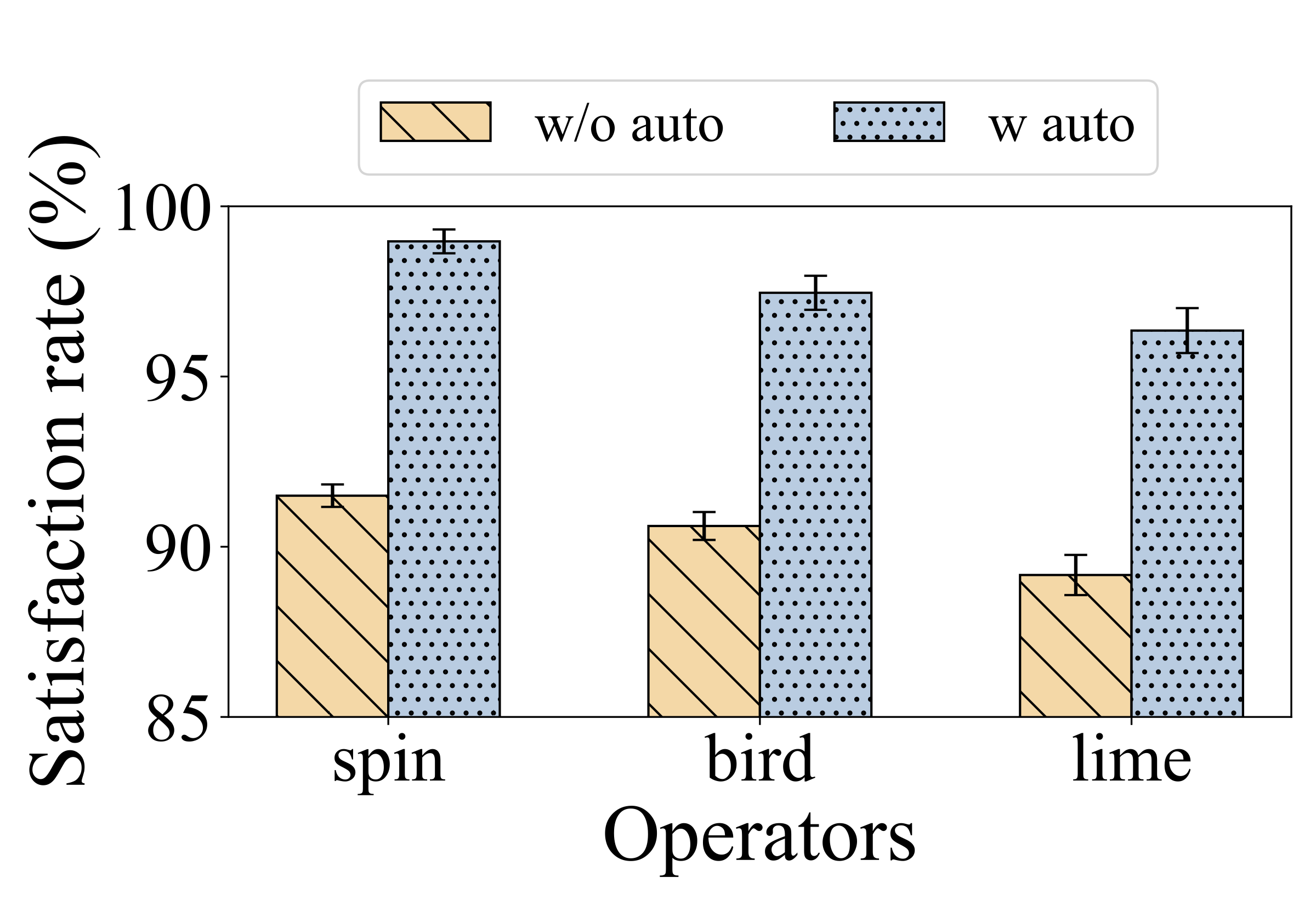}
    \vspace{-10pt}
    
    \caption{The effect of different system operators}
    \label{fig:vendor}
  \end{minipage}

  \label{fig:four_in_one_row}
\end{figure*}

\subsection{Ablation Study}

\subsubsection{\textbf{The effeciveness of vehicle redistribution}}
To demonstrate the effectiveness of the vehicle redistribution, we compare our method with the variant SMART w/o VRD, as shown in Table~\ref{tab:overall}. The experiment results show that removing the vehicle redistribution module results in a performance drop across all baselines, which indicates that strategic initialization based on predicted demand and post-scheduling traditional vehicle distribution plays a critical role in unlocking the full potential of ASMVs. Without vehicle redistribution, the autonomous vehicle scheduling model becomes similar to existing dynamic vehicle scheduling approaches~\cite{wang2021record,li2021dynamic}, which often overlook the importance of initial vehicle placement and rely solely on vehicle self-rebalancing, leading to suboptimal system performance. This, in turn, validates the importance of the vehicle redistribution module.

\subsubsection{\textbf{The effectiveness of vehicle self-rebalancing}}
To demonstrate the effectiveness of the vehicle self-rebalancing module, we compare our method with the variant SMART w/o VSR, as shown in Table~\ref{tab:overall}. The experiment results show that without the vehicle self-rebalancing module, ASMVs lose their ability to make high-frequency, time-sensitive self-rebalancing decisions based on spatial-temporally varied demand patterns. As a result, the autonomous scheduling model in this setting becomes functionally equivalent to reinforcement learning-based traditional vehicle scheduling models~\cite{tan2023joint,tan2024robust}, lacking the real-time responsiveness that characterizes autonomous systems. This underscores the importance of the vehicle self-rebalancing module.

\subsection{Impact of factors}
\subsubsection{\textbf{Different scales of ASMVs}}
To assess the effectiveness of autonomous vehicle scheduling under varying deployment scales, we evaluate the system performance across different ratios of autonomous vehicles relative to the total fleet. Figure~\ref{fig:ratio} demonstrates that even minimal deployments, such as 1 or 5 ASMVs (representing ratios of 0.04\% and 0.2\%, respectively), produce notable improvements in overall system efficiency. Specifically, the demand satisfaction rate increases from 90.61\% to 91.07\% with just one ASMV, and further to 92.14\% with five ASMVs. This underscores the value of autonomous scheduling even when fleet sizes are very small.
However, as the number of ASMVs increases, the marginal improvement in satisfaction rate gradually diminishes. For instance, increasing the fleet size from 77 to 155 vehicles (i.e., from 3$\%$ to 6$\%$ ratio) only yields a modest performance gain—from 97.84$\%$ to 98.76$\%$. Considering that the deployment cost of ASMVs is much higher than that of traditional non-electric vehicles~\cite{sanchez2020autonomous}, and that the performance improvement exhibits diminishing returns, we choose a ratio of 3$\%$ (i.e., 77 ASMVs) as a balanced and cost-effective deployment strategy for subsequent evaluations.

\label{sec:ratio}

\subsubsection{\textbf{Different fault rate}}
Considering that current autonomous driving technologies cannot yet guarantee a 100\% success rate in vehicle self-rebalancing (e.g., obstacles or accidents), we evaluate the robustness of our autonomous scheduling framework under fault conditions. Specifically, we simulate vehicle fault rates ranging from 0\% to 50\%, with faulty ASMVs being deactivated and excluded from operations.
As illustrated in Figure~\ref{fig:fault}, although system performance gradually declines as autonomous vehicle failure rates increase, the overall demand satisfaction rate remains relatively high, especially when compared to the 90\% baseline with no ASMVs involved (see Table~\ref{tab:overall}). This robustness primarily arises from our scheduling framework's capability to significantly enhance system performance even with a minimal deployment of ASMVs, inherently providing resilience under limited fleet availability. Additionally, the decision not to alter the existing scheduling of traditional vehicles, which continue to serve the majority of user requests, further contributes to maintaining system robustness.

\subsubsection{\textbf{Background Demand}}
In our main experiments, we adopt a common assumption in prior work~\cite{tan2023joint,tan2024robust,wang2021record} that recorded trips fully represent total user demand. However, this overlooks background demand—unobserved requests suppressed by insufficient vehicle availability. To address this, we estimate and introduce background demand to evaluate its impact on scheduling performance.
Our estimation leverages cumulative vehicle inflow/outflow and recorded demand over time. Specifically, if a region has negative cumulative net inflow and no recorded demand in a given period, we infer the presence of unmet background demand. For each such region-hour pair, we sample the number and energy cost of synthetic trips based on historical distributions observed in similar conditions. This approach preserves realistic spatiotemporal and energy usage patterns, enabling a more accurate assessment of background demand impacts, as illustrated in Figure~\ref{fig:sampling}.

The generated background trips account for approximately 7.85$\%$ of the total recorded trips. Figure~\ref{fig:sampling} illustrates the total demand satisfaction rate under varying autonomous vehicle ratios, before and after incorporating background demand.
We observe that introducing background demand leads to a decline in system performance across all settings. However, this performance gap diminishes as the proportion of ASMVs increases and becomes negligible at higher levels.
These results underscore the importance of considering background demand for accurate performance evaluation, and further demonstrate the robustness of autonomous vehicle scheduling to demand fluctuations.

\label{subsec:demand}

\subsubsection{\textbf{Multi-vendor demand}}
We assess the effectiveness of autonomous vehicle scheduling across different shared micromobility system operators, each characterized by distinct fleet sizes (Spin: 2,581 vehicles; Lime: 2,695 vehicles; Bird: 2,795 vehicles) and varying demand volumes (Spin: 181,019 trips; Lime: 169,491 trips; Bird: 278,665 trips). As illustrated in Figure~\ref{fig:vendor}, integrating autonomous vehicle scheduling consistently enhances the total demand satisfaction rate for all three operators, yielding improvements of at least 7.56\%. This result highlights the generalizability and effectiveness of our proposed approach in diverse operational settings.

\subsubsection{\textbf{The impact of integrating ASMVs on existing scheduling strategies}} 
\begin{figure}[t]\centering

\begin{minipage}[t]{0.45\linewidth}
    \includegraphics[width=\linewidth, keepaspectratio=true]{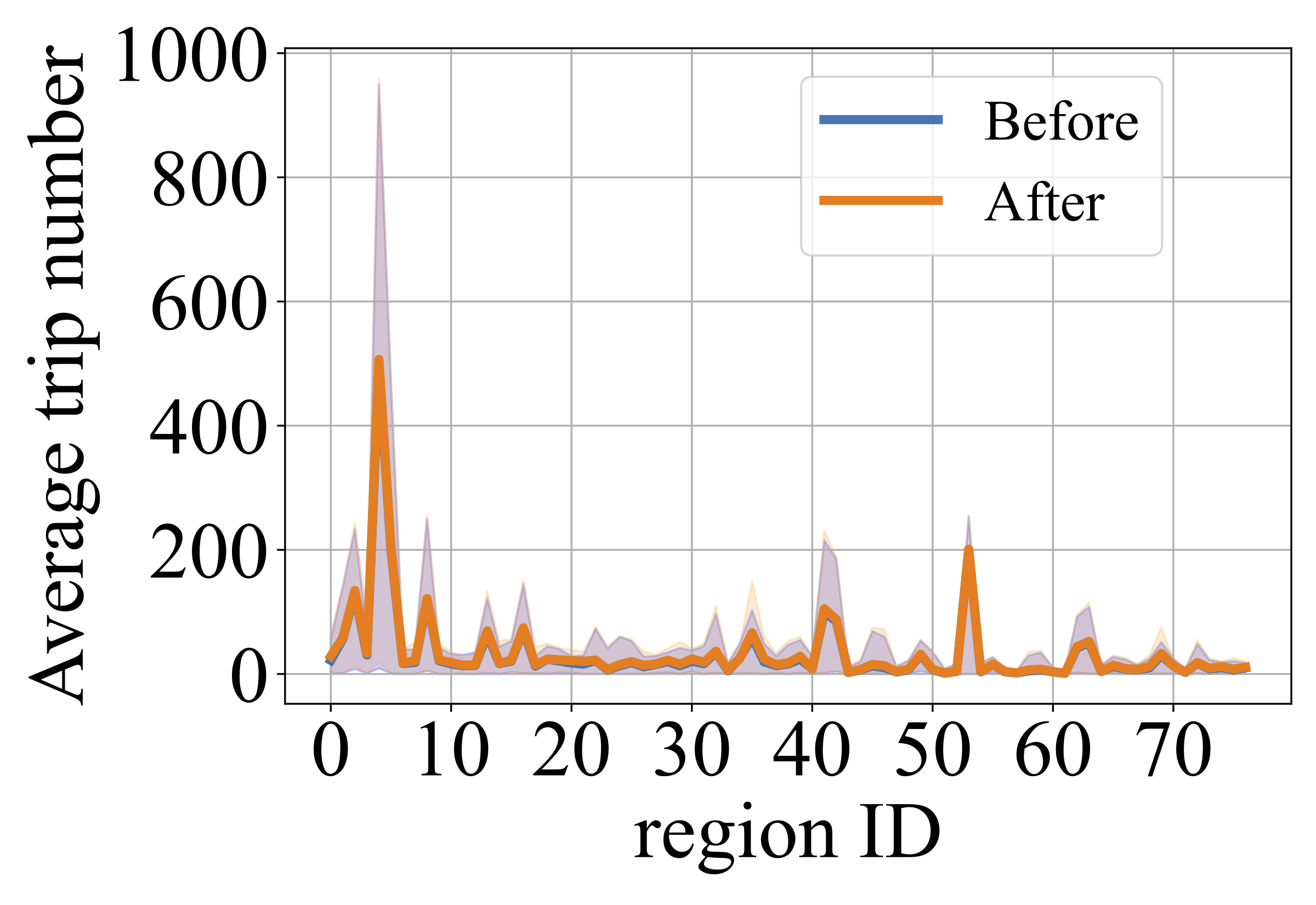}
    \vspace{-15pt}
    \captionsetup{font={small}}
    \caption{The satisfied demand distributions in different regions before and after integrating ASMVs}
    \label{fig:case2}
\end{minipage}
\hspace{10pt}
\begin{minipage}[t]{0.45\linewidth}
    \includegraphics[width=\linewidth, keepaspectratio=true]{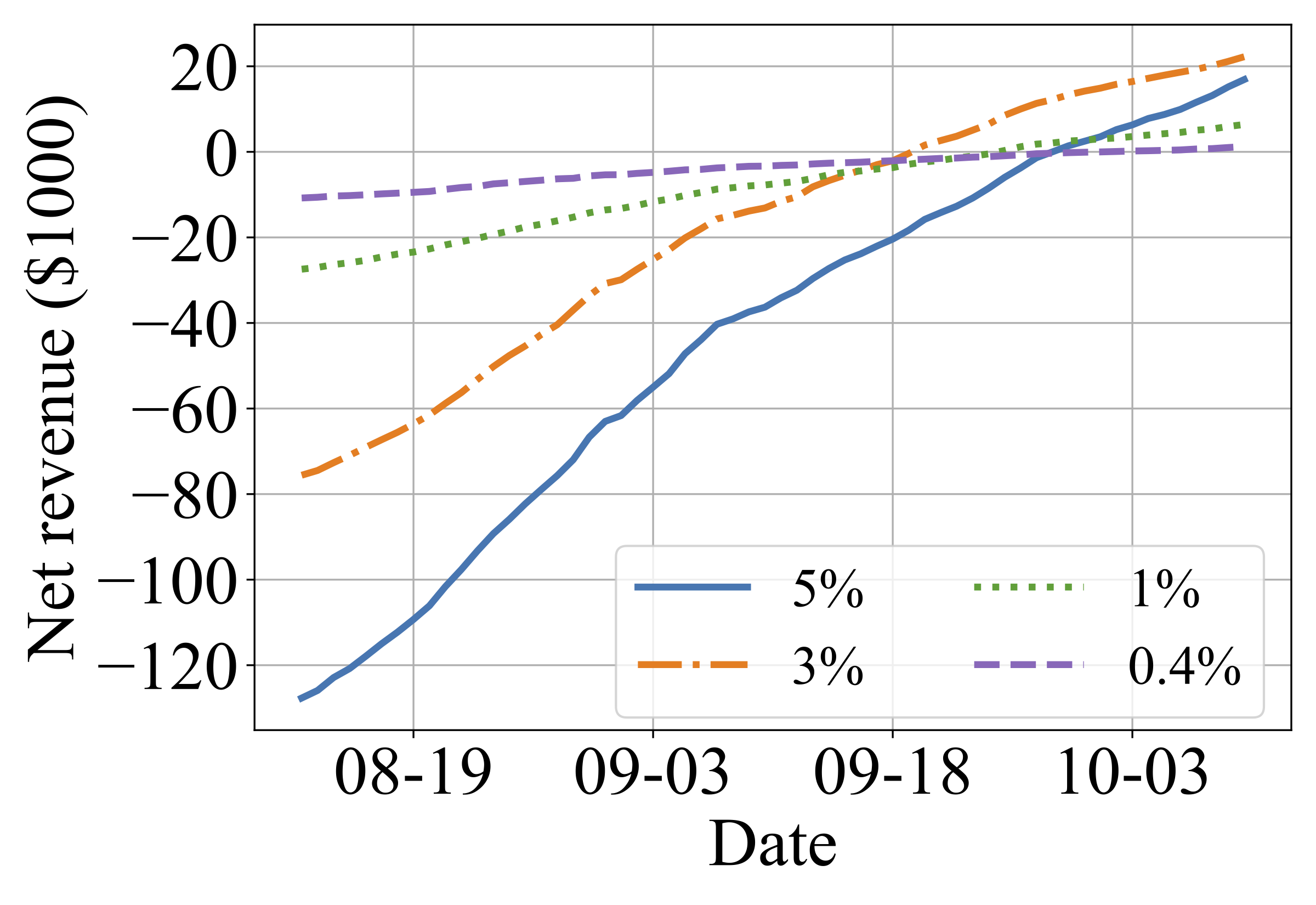}
    \vspace{-15pt}
    \captionsetup{font={small}}
    \caption{The net revenue of the system under different ASMV scales}
    \label{fig:case_study}
\end{minipage}
\vspace{-15pt}
\end{figure}

In our framework, the redistribution strategies for traditional shared micromobility vehicles are learned based on historical user demand patterns. 
Integrating ASMVs into the existing system enhances overall performance by satisfying additional demand, potentially causing a shift in demand distribution. Such a shift poses a risk of reducing the effectiveness of the traditional vehicles' schedules.
To evaluate this concern, we examine the demand distributions before and after incorporating ASMVs. Specifically, we analyze the average satisfied demand distributions across all regions over two months, comparing conditions with and without ASMVs, as depicted in Figure~\ref{fig:case2}. We apply the two-sample Kolmogorov-Smirnov (KS) test to these distributions (each consisting of 77 regional samples). The test results (KS statistic = 0.1509, p-value = 0.5865) indicate no statistically significant difference between the two distributions. This finding suggests that integrating ASMVs does not significantly alter the demand patterns and therefore does not undermine the existing strategies.

\subsection{Case Study}

To further evaluate the economic viability of integrating ASMVs into existing systems, we conduct a two-month case study analyzing the net revenue of ASMVs under different scales. Here, the net revenue is defined as the total trip revenue from satisfied demand minus autonomous vehicle deployment and charging costs. Specifically, the trip revenue of each user demand is calculated based on the trip duration and the operator's pricing scheme~\cite{chicago_price}. The deployment cost of an autonomous shared e-scooter is assumed as $\$$1000~\cite{sanchez2020autonomous,isinwheel2024}. As shown in Figure~\ref{fig:case_study}, we track the cumulative net revenue over time, accounting for trip revenue from satisfied user demand, charging costs, and deployment costs associated with ASMVs. Through this figure, we can know that while increasing the number of ASMVs generally accelerates net revenue accumulation, the time to break even does not decrease monotonically with scale. Specifically, the 3$\%$ ratio (77 ASMVs) achieves the earliest break-even point, reaching positive net revenue faster than both lower and higher deployment ratios, suggesting that the 3$\%$ setting achieves the best balance between operational gains and investment costs. The experiment results highlight a nonlinear relationship between ASMV fleet size and economic return: larger fleets not only enhance operational performance but also introduce heavier financial burdens at the early stage. Thus, careful tuning of the deployment scale is crucial for maximizing return on investment within a practical timeframe.

%% file: 2RelatedWork.tex
\section{Related Work}

\noindent \textbf{Shared Micromobility Vehicle Scheduling}: A considerable body of research has explored rebalancing strategies for shared micromobility vehicles~\cite{lee2024battery,yun2022automated,tan2025realism,tan2023joint,zhang2022multi,zhong2022bike,tan2023joint1,pan2019deep}, and those works can be divided into two categories based on their methodologies: (1) Some of them utilize optimization-based methods for shared micromobility vehicle scheduling: For example, \cite{lee2024battery} proposes a mixed integer programming (MIP) model for vehicle rebalancing and battery replacement in electric scooter systems, combined with personnel path selection to reduce operating costs. \cite{yun2022automated} proposes an optimized relocation scheme based on a genetic algorithm (GA) for shared electric scooters, which matches supply and demand through complete redistribution, optimizes operational efficiency, and reduces excess vehicle deployment. (2) Others utilize RL-based methods to find optimal scheduling policies: for example, \cite{tan2023joint} views each region as an agent and designs a MARL-based shared micromobility vehicle rebalancing and charging framework considering the energy consumption of user requests. \cite{zhang2022multi} utilizes a GCN-based MARL framework to solve the charging station request-specific dynamic pricing problem. They regard each charging station as an agent and consider the competitive-cooperative relationships between different agents to maximize operator benefits through vehicle scheduling. However, those methods focus on the scheduling problem for traditional shared micromobility vehicles, which follow an infrequent vehicle scheduling scheme, leading to an inherent bottleneck, primarily influenced by the significant spatio-temporal variability of user demand. Therefore they often fail to deliver stable performance under atypical conditions (e.g., the demand surge).

\noindent \textbf{Autonomous Shared Micromobility Vehicles}: Recent studies have explored the potential of autonomous shared micromobility vehicles (ASMVs) from various perspectives~\cite{wu2025towards,sanchez2020autonomous,sanchez2020autonomous,kondor2021estimating,coretti2023urban}. For example, \cite{sanchez2020autonomous} introduces the concept of ASMVs and presents a functional prototype capable of self-navigation and stability control. \cite{wu2025towards} design a city-scale simulation platform to evaluate the navigation models of ASMVs under realistic urban settings. \cite{sanchez2024shared} outlines an implementation roadmap that positions ASMV as a core component of sustainable and walkable urban mobility systems. Different from these works, which focus primarily on the feasibility and prototyping of ASMVs, our work assumes the availability of such capabilities and explores how to integrate them into an existing shared micromobility system to augment the overall vehicle scheduling efficiency.

%% file: 7Conclusion.tex
\section{Conclusion}
In this work, we focus on the problem of integrating ASMV scheduling into traditional vehicle scheduling. We design a hierarchical reinforcement learning framework called SMART, which incorporates both ASMV redistribution and vehicle self-rebalancing. The evaluation results show that SMART achieves the most significant performance improvement over all three traditional vehicle scheduling baselines, achieving an improvement of at least 7.56$\%$ in demand satisfaction rate.